\documentstyle[pra,aps,eqsecnum]{revtex}
\begin{document}
\title
{
From quantum Bayesian inference to quantum tomography
\thanks{We dedicate this paper to the sixtieth birthday of Professor
Jan Pe\v{r}ina.}
}

\author{
R. Derka$^{1}$, V. Bu\v{z}ek$^{1,2}$, G. Adam$^{3}$, and P.L. Knight$^{2}$
}
\address{$^{1}$ Institute of Physics, Slovak Academy of Sciences,
D\'ubravsk\'a cesta 9, 842 28 Bratislava, Slovakia\\
$^{2}$ Optics Section, The Blackett Laboratory,
Imperial College, London SW7 2BZ,  England\\
$^{3}$ Institut f\"ur Theoretische Physik, Technische Universit\"at Wien,\\
    Wiedner Hauptstrasse 8-10, A-1040 Vienna, Austria
}

\date{December 13, 1996}
\maketitle

\begin{abstract}
We derive an expression for a density operator estimated via
Bayesian quantum inference  in the limit of an infinite
number of measurements.
 This expression is derived under the
assumption that the reconstructed system is in a pure state.
In this case the estimation corresponds
to an averaging over a generalized microcanonical ensemble of pure states
satisfying a set of  constraints imposed by the measured mean values of
the observables under consideration.
We show that via the ``purification'' ansatz, statistical mixtures
can also be consistently reconstructed via the quantum Bayesian
inference scheme.
In this case the estimation corresponds to averaging over the
generalized canonical ensemble of states satisfying the given constraints,
and the reconstructed density operator maximizes the von\,Neumann entropy
(i.e., this density operator is equal to the generalized canonical density
operator which follows from the Jaynes principle of
maximum entropy).
We study in detail the reconstruction of the spin-1/2 density operator and
discuss the logical connection between the three reconstruction schemes, i.e.,
{\bf (1)} quantum Bayesian inference, {\bf (2)}  reconstruction
via the Jaynes principle of maximum entropy, and {\bf (3)} discrete
quantum tomography.
\end{abstract}

\section{INTRODUCTION}

The essence of the problem of  state determination
($\equiv$ state reconstruction) lies in an {\em a posteriori}
estimation of a density operator ($\equiv$ corresponding quasiprobability
density distribution) of a quantum-mechanical (microscopic) system
based on data obtained with the help of a macroscopic measurement
apparatus
\cite{Ballentine90}.
The quality of the reconstruction depends on the quality of
the measured data and the efficiency of the reconstruction procedure
with the help of which the inversion-data analysis is performed.
In particular, we can specify three different situations: Firstly,
when all system observables are precisely measured;
Secondly, when just part of the system observables is precisely
measured; Finally,  when measurement does not provide
us with information enough to specify exact mean values (or probability
distributions) of observables under consideration.

\subsection{Complete observation level}
Providing all system observables (i.e., the quorum
\cite{Band79})
have  been precisely measured, then the density operator
of a quantum-mechanical
system can be completely reconstructed (i.e., the density operator
can be uniquely determined based on the available data).
In principle,
we can consider two different schemes for reconstruction of the
density operator (or, equivalently, the Wigner function)
of the given quantum-mechanical  system. The
difference between these two schemes is based on the way in which
information about the quantum-mechanical system is obtained. The first
type of  measurement is such that on each element of the ensemble
of the measured states
only a {\em single} observable is measured.
In the second type of measurement
 a {\em simultaneous} measurement of conjugated
observables is assumed. We note that
in both cases we assume an ideal, i.e.,  unit-efficiency,
measurements.

\subsubsection{Quantum tomography}
When the single-observable measurement is performed, a {\em distribution}
$W_{|\Psi\rangle}(A)$ for a particular observable $\hat{A}$ in the state
$| \psi\rangle$ is obtained in an unbiased way
\cite{Neumann55},
 i.e., $W_{|\Psi\rangle}(A)=
|\langle\Phi_A| \Psi\rangle|^2$, where $| \Phi_A\rangle$
are eigenstates of the observable $\hat{A}$ such that
$\sum_A|\Phi_A \rangle\langle\Phi_A |=\hat{1}$.  Here a question arises:
What is the {\em smallest} number of distributions $W_{|\Psi\rangle}(A)$
required to determine the state uniquely? If we consider the reconstruction
of the state of a harmonic oscillator, then
this question is directly
related to the so-called Pauli problem
\cite{Pauli80}
 of the reconstruction of the wave-function from
distributions $W_{|\Psi\rangle}(q)$ and $W_{|\Psi\rangle}(p)$ for the
position and
momentum of the state $| \Psi\rangle$. As shown by Gale, Guth and Trammel
\cite{Gale68}
for example, the knowledge of $W_{|\Psi\rangle}(q)$ and $W_{|\Psi\rangle}(p)$
is not in general sufficient for a complete reconstruction of the wave
(or, equivalently, the Wigner)
function. In contrast, one can consider an {\em infinite}
set of distributions $W_{|\Psi\rangle}(x_{\theta})$ of the rotated quadrature
$\hat{x}_{\theta}=\hat{q}\cos\theta +\hat{p}\sin\theta$. Each distribution
$W_{|\Psi\rangle}(x_{\theta})$	can be obtained in a measurement of a {\em
single}
observable $\hat{x}_{\theta}$, in which case a detector (filter) is prepared
in an eigenstate $|x_{\theta} \rangle$ of this observable. It has been
shown by Vogel and Risken
\cite{Vogel89}
that from an infinite set (in the case of the harmonic oscillator)
of the measured distributions $W_{|\Psi\rangle}(x_{\theta})$
for all values of $\theta$ such that $[0<\theta\leq\pi]$, the
Wigner function can be reconstructed uniquely via the inverse Radon
transformation. In other words knowledge of the set of distributions
$W_{|\Psi\rangle}(x_{\theta})$ is equivalent to  knowledge of the Wigner
function
(or, equivalently, the density operator).
This scheme for reconstruction of the Wigner function (the so called
{\em optical homodyne tomography}) has recently been realized experimentally
by Raymer and his coworkers \cite{Smithey93}. In these experiments
Wigner functions of
a coherent state and a squeezed vacuum state have been
reconstructed from tomographic data. Quantum-state tomography can be
applied not only to optical fields (harmonic oscillators) but
for  reconstruction of other physical systems, such as
atomic waves (see recent work by Janicke and Wilkens  \cite{Wilkens}).
Leonhardt \cite{Leon95b} has recently developed a theory of  quantum
tomography of discrete Wigner functions describing states of quantum
systems with  finite-dimensional Hilbert spaces (i.e.,
angular momentum or  spin).

\subsubsection{Filtering with quantum rulers}

In the case of the simultaneous measurement of two non-commuting
observables (let us say $\hat{q}$ and $\hat{p}$) it is not possible
to construct a joint eigenstate of these two operators, and therefore it is
inevitable that the simultaneous measurement of two non-commuting observables
 introduces additional
noise (of quantum origin) into measured data. This noise is associated
with Heisenberg's  uncertainty relation and it results in
a specific ``smoothing'' (equivalent to a reduction of resolution)
of the original  Wigner function of the system under consideration
(see Refs. \cite{Husimi} and \cite{Arthurs65}).
To describe a process of simultaneous measurement of two non-commuting
observables, W\'odkiewicz
\cite{Wod84}
has proposed a formalism
based on an operational probability density distribution which explicitly
takes into account the action of the measurement device modelled as a
``filter'
(quantum ruler). A particular choice of the state of the ruler samples a
specific  type of accessible information concerning the system, i.e.,
information about the system is biased by the filtering process. The
quantum-mechanical noise induced by filtering formally results in a
smoothing
of the original Wigner function of the measured state
\cite{Husimi,Arthurs65},
so that the operational probability
density distribution can be expressed as a convolution of the original
Wigner function and the  Wigner function of the filter state.
In particular, if the filter is considered to be in
its  vacuum state then the corresponding operational probability density
distributions is equal to the Husimi ($Q$) function \cite{Husimi}.
The $Q$ function of optical fields has been experimentally
measured using such an approach
by Walker and Carroll \cite{Walker86}. The direct experimental
measurement of the operational probability density distribution
with the filter in an arbitrary state is feasible in an 8-port
experimental  setup of the type
used by Noh, Foug\'eres and Mandel \cite{Noh91} (for more details,
see the recent book by Pe\v{r}ina and coworkers \cite{Perina}).

As a consequence of a simultaneous measurement of non-commuting
observables
the measured distributions
are fuzzy (i.e., they are equal to smoothed Wigner functions). Nevertheless,
if the detectors used in the experiment have a unit efficiency (in the
case of an ideal measurement), the noise induced by quantum filtering can be
``separated'' from the measured data and the density operator (Wigner
function) of the measured system
can be
``extracted'' from the operational probability density distribution. In
particular, the Wigner function can be uniquely reconstructed from the
$Q$ function. This extraction procedure is
technically quite involved and it suffers significantly if additional
stochastic noise due to imperfect measurement is present in the data.

We note that
propensities, and in particular $Q$-functions, can be also associated
with discrete phase space  and they can in principle be
measured directly \cite{Opat95}. These discrete probability distributions
contain complete information about density operators of measured
systems.

\subsection{Reduced observation levels and MaxEnt principle}

As we have already indicated it is now well understood that density operators
(or Wigner functions)
can, in principle, be uniquely reconstructed using either the single
observable measurements (optical homodyne tomography) or the simultaneous
measurement of two non-commuting observables. The completely reconstructed
density operator (or, equivalently, the Wigner function)
 contains information about {\em all} independent moments
of the system operators. For example, in the case of the quantum harmonic
oscillator,
the knowledge of the Wigner function is equivalent to the knowledge of all
moments $\langle(\hat{a}^{\dagger})^m\hat{a}^n\rangle$ of the creation
$(\hat{a}^{\dagger})$ and annihilation $(\hat{a})$ operators.

In many cases it turns out that the state of a harmonic oscillator
 is characterized
by an {\em infinite} number of independent moments
$\langle(\hat{a}^{\dagger})^m\hat{a}^n\rangle$ (for all $m$ and $n$).
Analogously, the state of a quantum system in a finite-dimensional
Hilbert space can be characterized by a very large number
of independent parameters.
A {\em complete} measurement of these moments can  take an
infinite time to perform. This means that even though the Wigner function
can in principle
be reconstructed the collection of a complete set of experimental data points
is (in principle) a  never ending process.
In addition the data processing and numerical reconstruction of the Wigner
function are time consuming. Therefore experimental realization of
the reconstruction of the density operators (Wigner functions) for many
systems   can be problematic.

In practice, it is possible to perform a measurement of just a finite number
of independent moments of the system operators which means that only
a subset $\hat{G}_{\nu}$ ($\nu=1,2,...,n)$ of observables from
the quorum (this subset constitutes the so-called observation level
\cite{Jaynes57})
is measured. In this case, when the complete information about the
system is not available, one needs an additional criterion which would
help to reconstruct (or estimate) the density operator uniquely.
Provided mean values of all observables on the given observation
level are measured precisely, then the density operator
(the Wigner function) of the system under consideration can be
reconstructed with the help of the Jaynes principle of maximum
entropy (the so called {\em MaxEnt} principle)
\cite{Jaynes57}.
The reconstructed density operator fulfills several conditions.
Firstly, its trace has to be equal to unity (i.e., ${\rm Tr}\hat\rho =1$).
Secondly,
${\rm Tr}(\hat\rho\hat{G}_{\nu})=G_{\nu}$ ($\nu=1,2,...,n)$
which means that the
reconstructed density operator provides us with the measured mean values
of those observables which constitute the given observation level.
Obviously, a large number of density operators can fulfill these
two constraints. So one needs an additional criterion which
would uniquely specify the generalized canonical density operator.
According to Jaynes \cite{Jaynes57}
this operator has to be that one with the largest
value of the von\,Neumann entropy $S=-{\rm Tr}\{\hat\rho\ln\hat\rho\}$.
This additional condition means that the {\em MaxEnt} principle is the
most conservative assignment in the sense that it does not permit
one to draw any conclusion not warranted by the experimental data
\cite{comment}.

The MaxEnt principle provides us with a very efficient
prescription how to reconstruct density operators of quantum-mechanical
systems providing mean values of a given set of observables are known.
It works perfectly well for systems with (semi)infinite Hilbert spaces
(such as quantum-mechanical harmonic oscillator) as well as for
systems with finite-dimensional Hilbert spaces (such as spin systems).
If the observation level is composed of the quorum of the observables
(i.e. this is a complete observation level), then the
MaxEnt principle represents an alternative to  quantum tomography,
i.e. both schemes are equally suitable for the analysis of the
tomographic data (for details see \cite{Buzek96}). To be specific,
the observation level in this case is composed of all projectors
associated with probability distributions of rotated quadratures.
The power of the MaxEnt principle can be appreciated in analysis
of incomplete tomographic data (equivalent to a reconstruction
of the Wigner function in the discrete phase space).
In particular, Wiedemann \cite{Wiedemann96} has performed a
numerical reconstruction of the density operator (Wigner function)
from incomplete tomographic data based on the MaxEnt principle
as discussed by Bu\v{z}ek et al. \cite{Buzek96}.   Wiedemann has shown
that in particular cases {\em MaxEnt} reconstruction  from incomplete
tomographic data can be  several
orders better compared to a standard tomographic inversion.
These results can be interpreted as suggesting that
 the MaxEnt principle is the conceptual
basis behind the tomographic reconstruction (irrespective whether
in continuous or discrete phase spaces).

\subsection{Incomplete measurement and Bayesian inference}

It has to be stressed that the Jaynes principle of  maximum
entropy can consistently be applied only when {\em exact}
mean values of the measured observables are available. This
condition implicitly assumes that an infinite number of repeated measurements
on different elements of the ensemble
has to be performed to reveal the exact mean value of the given observable.
In practice only a finite number of measurements can be performed.
What is obtained from these measurements is a specific set
of data indicating number of how many times eigenvalues of given observables
have appeared (which in the limit of an infinite number of measurements
results in the corresponding  quantum probability distributions).
The question is, how to obtain the best {\em a posteriori} estimation
of the density operator based on the measured data.
Helstrom \cite{Helstrom76},
Holevo \cite{Holevo82},
 and Jones \cite{Jones91} have shown that the answer to this
question can be given by the Bayesian inference method, providing it is
{\em a priori} known that the quantum-mechanical state which is going
to be reconstructed is prepared in a pure (even though unknown) state.
Once this purity condition is fulfilled, then the observer can systematically
estimate (i.e. reconstruct) an {\em a posteriori} density operator
based upon an incomplete set of experimental data. This density operator
is equal to the mean over all possible {\em pure} states weighted by
a specific probability distribution in an abstract state space with the
unique invariant integration measure. It is this probability distribution
(conditioned by the assumed Bayesian prior) which characterizes observer's
knowledge of the system at every moment during the measurement sequence.
We note once again that the Bayesian  inference has been developed
for a reconstruction of {\em pure} quantum mechanical states and in this
sense it corresponds to an averaging over a {\em generalized microcanonical}
ensemble.

In a real situation one can never design a state-preparation device
 such that it produces an
ensemble of identical pure states. What usually happens is that the ensemble
consists of a set of pure states, each of which is represented in the
ensemble with a certain probability (alternatively, we can say that
the system under consideration is entangled with other quantum-mechanical
systems).
So now the question is how to use
the Bayesian reconstruction scheme when the quantum-mechanical
system under consideration is in  an impure state (i.e., a statistical
mixture). To apply the Bayesian inference scheme, one has to define exactly
three objects: {\bf (1)} the abstract state space of the measured
system; {\bf (2)} the corresponding invariant integration measure
of this space; and {\bf (3)} the {\em prior} (i.e., the {\em a priori} known
probability distribution on the given parametric state space).
Once these objects are specified one can estimate an {\em a posteror}
density operator after each individual outcome of the measurement has been
registered.

The main purpose of the present paper is to show from an example of the
state-reconstruction of a spin-1/2 system
how the Bayesian  scheme of quantum inference
developed for a reconstruction of statistical mixtures \cite{Derka96}
actually works. We will show that this scheme
corresponds to a specific averaging over the grand canonical ensemble.
Moreover,
we will show that
in the limit of infinite number of measurements the reconstructed
density operator is {\em equal} to the generalized canonical
density operator obtained via the Jaynes principle of  maximum entropy.
In addition, in the case the complete observation level
(the quorum of the observables is measured) the generalized canonical
density operator is equal to the operator obtained via the tomographic
measurement. This clearly reveals a logical connection between
quantum Bayesian inference and	quantum tomography.

The paper is organized as follows. In Section II we briefly review the
Bayesian inference scheme as developed for pure states by Jones
\cite{Jones91}. In Section III we derive limiting formula for
an {\em a posteriori} estimated density operator of any quantum
mechanical system under the prior assumption that the system is in
a pure state.  Section IV is devoted to a description of Bayesian
inference for statistical mixtures. Reconstructions of a
density operator of a spin-1/2 are presented in Section V.
We will conclude our paper with general remarks.

\section{BAYESIAN INFERENCE}

The general idea
of the Bayesian reconstruction scheme is based on manipulations with
probability
distributions in parametric state spaces. To understand this reconstruction
scheme we remind us several definitions and concepts. Firstly, it is
a space of states of the measured system.
The quantum Bayesian method  as discussed in the literature
\cite{Helstrom76,Holevo82,Jones91}
is based on the assumption
that the reconstructed system is in a pure
state described by a state vector $|\Psi\rangle$,
or equivalently by a pure-state density
operator $\hat{\rho}=|\Psi\rangle\langle\Psi|$.
The manifold of all pure states is a continuum which we denote as
$\Omega$. Secondly,  it is the discrete space $A$ of reading states
of a measuring	apparatus associated with the observable $\hat{O}$.
These states are intrinsically related to the projectors
$\hat{P}_{{\lambda_i,\hat O}}$, where
$\lambda_i$ are the eigenvalues of the observable $\hat O$.

The Bayesian reconstruction scheme
is formulated as a three-step inversion procedure:\newline
{\bf (1)} As a result of a measurement a conditional probability
\begin{eqnarray}
p(\hat O,\lambda_i \vert \hat{\rho})={\rm Tr}
\left(\hat{P}_{\lambda_i,\hat O}\hat{\rho}\right),
\label{3.1}
\end{eqnarray}
on the discrete space $A$ is defined.
This conditional probability distribution  specifies a probability of finding
the result $\lambda_i$ if the measured system is in a particular
state $\hat{\rho}$.\newline
{\bf (2)} To perform the second step of the inversion procedure
one has to specify an {\em a priori} distribution
$p_0(\hat{\rho})$ defined on the space	$\Omega$. This distribution
describes our initial knowledge concerning the measured system. Using the
conditional probability distribution
$p(\hat O,\lambda_i \vert \hat{\rho})$ and the {\em a priori} distribution
$P_0(\hat{\rho})$ we can define the {\em joint} probability distribution
$p(\hat O,\lambda_i ; \hat{\rho})$
\begin{eqnarray}
p(\hat O,\lambda_i ; \hat{\rho})= p(\hat O,\lambda_i \vert \hat{\rho})
p_0(\hat{\rho}),
\label{3.2}
\end{eqnarray}
on the space $\Omega\otimes A$. We note that if no initial information
about the measured system is known, then the prior $p_0(\hat{\rho})$
has to be assumed to be constant (this assumption is related to the
Laplace principle of indifference \cite{comment}).\newline
{\bf (3)} The final step of the Bayesian reconstruction is based on the
well known Bayes rule $p(x|y)p(y)=p(x;y)=p(y|x)p(x)$, with the help
of which we find the conditional probability
$p(\hat{\rho} \vert \hat O,\lambda_i)$ on the state space $\Omega$:
\begin{eqnarray}
p(\hat{\rho} \vert \hat O,\lambda_i)=
{{p(\hat O,\lambda_i ,\hat{\rho})}\over
{\int _\Omega p(\hat O,\lambda_i ,\hat{\rho}) d_\Omega}},
\label{3.3}
\end{eqnarray}
from which the reconstructed density operator can be obtained
[see Eq.(\ref{3.4})].

In the case of the repeated $N$-trial measurement, the reconstruction
scheme consists of an
iterative utilization of the three-step procedure as described above.
 After the $N$-th measurement we
use as an input for the prior distribution the conditional
probability distribution which is an output after the $(N-1)$st measurement.
However, we can equivalently define the $N$-trial measurement conditional
probability
$p(\{\ \}_{_N}\vert \hat{\rho})=
\prod_{i=1}^N p(\hat O_i,\lambda_j \vert \hat{\rho})$ and
applying the three-step procedure
just once we get the
reconstructed density operator
\begin{eqnarray}
\hat{\rho}(\{ \ \}_{N})={\int _\Omega p(\{\ \}_{N}\vert \hat{\rho})
\hat{\rho} d_\Omega \over
 {\int _\Omega p(\{\ \}_{N}\vert \hat{\rho}) d_\Omega}},
\label{3.4}
\end{eqnarray}
where $\hat\rho$ in the r.h.s. of Eq.(\ref{3.4}) is a properly parameterized
density operator in the state space $\Omega$.
At this point we should mention one essential problem in the Bayesian
reconstruction scheme, which is the determination of the integration measure
$d_\Omega$ \footnote{
Many authors (see, for instance, Ref.\cite{Jones91})
identify the prior distribution
with the integration measure on the space $\Omega$. However, the
particular form
of $d_\Omega$ is associated with the topology and the
particular parameterization of
the space $\Omega$ rather then with some prior information
$p_0(\hat{\rho})$ about this system.
We will distinguish between these two objects.}.

The integration measure has to be invariant under
unitary transformations in the space $\Omega$. This
requirement uniquely determines the form of the measure.
However, this is no longer valid when
$\Omega $ is considered to be
a  space of mixed states formed by all convex
combinations of elements of the original pure state space
$\Omega$. Although the Bayesian procedure itself does not require any special
conditions imposed on the space $\Omega$, the ambiguity
in determination of the integration measure has been
the main obstacle in generalization of the Bayesian inference scheme
for a reconstruction of {\em a priori} impure quantum states.

\section{BAYESIAN INFERENCE IN LIMIT OF INFINITE NUMBER OF MEASUREMENTS}

The explicit evaluation of an {\em a posteriori} estimation of the
density operator $\hat{\rho}(\{\ \}_{N})$ is significantly limited
by technical difficulties when integration over parametric space
is performed [see Eq.(\ref{3.4})].
Even for the simplest quantum systems  and for	a relatively small number
 of measurements,
the reconstruction procedure can be technically insurmountable
 problem.

On the other hand let us assume that the number of measurements of
observables
$\hat O_i$
approaches  infinity (i.e.  $N\rightarrow \infty$).
It is clear that in this case
mean values of all projectors $\langle \hat{P}_{\lambda_j,\hat O_i}\rangle$
associated with the observables $\hat O_i$
are {\em precisely} know (measured): i.e.
\begin{eqnarray}
\langle \hat{P}_{\lambda_j,\hat O_i}\rangle=\alpha _j^i,
\label{4.1}
\end{eqnarray}
where $\sum_j \alpha_j^i=1$. In this case the integral in the
right-hand side of Eq.(\ref{3.4}) can be significantly
simplified  with the help of  the following lemma:
\medskip

\noindent{\it Lemma:} Let us define the integral expression
\begin{eqnarray}
I(\alpha_1,\dots,\alpha_{n-1})\equiv
\int_0^1 dx_1 \int_0^{y_2}dx_2\dots
\int_0^{y_{n-1}}
dx_{n-1}\,
F(x_1,\dots,x_{n-1}|\, \alpha_1,\dots,\alpha_{n-1})
\label{4.2}
\end{eqnarray}
where
\begin{eqnarray}
F(x_1,\dots,x_{n-1}|\, \alpha_1,\dots,\alpha_{n-1})
={1\over B}
x_1^{\alpha_1N}x_2^{\alpha_2N}\dots x_{n-1}^{\alpha_{n-1}N}
(1-x_1\dots -x_{n-1})^{\alpha_{n}N}
\label{4.3}
\end{eqnarray}
and $\alpha_j$ satisfy condition $\sum_j^n \alpha_j=1$. The integration
boundaries $y_k$ are given by relations:
\begin{eqnarray}
y_k = 1 -\sum_{j=1}^{k-1} x_j;\qquad k=2,\dots, n-1
\label{4.4}
\end{eqnarray}
and  $B$ equals to
the product of Beta functions $B(x,y)$:
\begin{eqnarray}
B \equiv B (a_{n}+1,a_{n-1}+1)
B (a_{n}+a_{n-1}+1,a_{n-2}+2)\dots
B (a_{n}+a_{n-1}\dots a_2+1,a_1+n-1).
\label{4.5}
\end{eqnarray}

{\it i.} \noindent
The function $F(x_1,\dots,x_{n-1}|\, \alpha_1,\dots,\alpha_{n-1})$
in the integral (\ref{4.2}) is a normalized probability
distribution in the $(n-1)$-dimensional volume given by integration
boundaries.

\noindent
{\it ii.}
For $N\rightarrow \infty$,
this probability distribution has the
following properties:
\begin{eqnarray}
\langle x_i\rangle \rightarrow \alpha_i\qquad
\langle x_i^2\rangle \rightarrow \alpha_i^2\qquad
i=1,2,3,\dots,n-1,
\label{4.6}
\end{eqnarray}
i.e., this probability density tends to the product of delta
functions:
\begin{eqnarray}
\lim_{N\rightarrow\infty}
F(x_1,\dots,x_{n-1}|\, \alpha_1,\dots,\alpha_{n-1})=
\delta (x_1- \alpha_1)\delta (x_2-\alpha_2)\dots \delta
(x_{n-1}-\alpha_{n-1}).
\label{4.7}
\end{eqnarray}

{\it Proof: i.} can be derived by a successive application
of an equation [see for example \cite{Gradstein}, Eqs.(3.191)]
\begin{eqnarray}
\int_0^u x^{\nu-1}(u-x)^{\mu-1}dx=u^{\mu +\nu -1}B(\mu, \nu ).
\label{4.8}
\end{eqnarray}

\noindent
{\it ii.} can be obtained as a result  of a straightforward calculations
of limits of certain
expressions containing Beta functions with integer-number arguments.
In our calculations we have used
the identity
\begin{eqnarray}
\frac{B(n+1,m)}{B(n,m)}=\frac{n}{n+m},
\label{4.9}
\end{eqnarray}
which is satisfied by Beta functions with integer-number arguments.

\subsection{Conditional density distribution}
Let us start with the expression for  conditional  probability distribution
$p(\{\ \}_{_{N}}\vert \hat\rho )$
for the $N$-trial measurement of a set of observables $\hat O_i$.
If we assume, that the number of measurements of each observable
$\hat O_i$ goes to the infinity then we can write:
\begin{eqnarray}
p(\{\ \}_{_{N\rightarrow\infty}}\vert \hat\rho )=\lim_{N\rightarrow\infty}
\prod_i \Big\lbrack
\prod_{j=1}^{n_i} {\rm Tr}\left(\hat P_{_{\lambda_j,\hat O_i}}
\hat\rho\right)^{\alpha _j^iN}\Big\rbrack.
\label{4.10}
\end{eqnarray}
The first product in the right-hand side (r.h.s.)
 of Eq.(\ref{4.10})
is associated with each measured observable $\hat O_i$ on
a given  observation level.
The second  product
runs over   eigenvalues $n_i$ of each observable $\hat O_i$.

In what follows we formally rewrite the  r.h.s. of Eq.(\ref{4.10}):
we insert in it a set of $\delta$-function and we perform the following
 integration
\begin{eqnarray}\nonumber
p(\{\ \}_{_{N\rightarrow\infty}}\vert \hat\rho )=\prod_i \left\{
\int_0^1dx^i_1 \int_0^{y_{2}^i} dx^i_2	   \dots
\int_0^{y_{n_i-1}^i}
dx^i_{n_i-1}\right.
 \delta\left[x^i_1-{\rm Tr}\left(\hat P_{_{\lambda_1,\hat O_i}}
\hat \rho\right)
\right]\dots
\end{eqnarray}
\begin{eqnarray}
\times\left.
\delta\left[x^i_{n_i-1}-{\rm Tr}\left(P_{_{\lambda_{n_i-1},\hat O_i}}
\hat{\rho}\right)
\right]
\prod_{j=1}^{n_i-1}(x^i_j)^{\alpha _j^iN}\ \ \ (1-x_1^i\dots x^i_{n_i-1})^
{\alpha^i_{n_i}N}\right\}
\label{4.11}
\end{eqnarray}
In Eq.(\ref{4.11}) we perform an integration over a volume
determined by the integration boundaries $y_k^i$ [see
Eq.(\ref{4.4})], i.e., due to the condition
$\sum^{n_i}_{j=1}{\rm Tr}(\hat{P}_{\lambda_j,\hat O_i}\hat{\rho})=1$,
there is no need to perform integration from $-\infty$ to $\infty$.

At this point we utilize our {\em Lemma}. To be specific,  firstly
we separate in	Eq.(\ref{4.11}) the term, which corresponds to the
 function $I$ given by Eq.(\ref{4.2}). Then we replace this term by
its limit expression (\ref{4.7}). After a
straightforward integration over variables $x^i_j$
we finally obtain an explicit expression for the conditional
probability
$p(\{\ \}_{_{N\rightarrow\infty}}\vert \hat\rho )$
which we insert into Eq.(\ref{3.4}) from which we obtain the  expression
for  an {\em a posteriori} estimation of the density operator
$\hat\rho(\{\ \}_{N\rightarrow \infty})$ on the given observation level:
\begin{eqnarray}
\hat{\rho}(\{\ \}_{_{N\rightarrow \infty}})={1 \over {\cal N}}
\int_\Omega \prod_i \left\{
\prod_{j=1}^{n_i-1}\delta \left[{\rm Tr}
\left(\hat{P}_{\lambda_j,\hat O_i}\hat{\rho}\right)-\alpha _j^i\right]
\right\}
\hat{\rho} d_\Omega.
\label{4.12}
\end{eqnarray}
Here ${\cal N}$ is a normalization constant determined
by the condition
${\rm Tr}\left[\hat{\rho}(\{\ \}_{_{N\rightarrow \infty}})\right]=1$.

The interpretation of Eq.(\ref{4.12}) is straightforward.
The reconstructed density operator is equal to the sum of
equally-weighted
pure-state density operators on the manifold $\Omega $, which do satisfy
the conditions given by Eq.(\ref{4.1})
[these conditions
are guaranteed by the presence of $\delta$-functions in
the r.h.s. of Eq.(\ref{4.12})].
In terms of statistical physics Eq.(\ref{4.12}) can be interpreted as
 an averaging over the generalized microcanonical
ensemble of those pure states which do
satisfy the conditions on the mean values of
the measured observables. Consequently, Eq.(\ref{4.12}) represents the
principle of the ``maximum entropy''
 on the generalized microcanonical ensemble under the
constraint (\ref{4.1}).

In order to clarify the relationship between the reconstruction procedure
based on the Jaynes principle of maximum entropy and
the quantum Bayesian inference
we present in Section V an
 example of the reconstruction of the
state of a  spin-1/2  system.

\section{BAYESIAN RECONSTRUCTION OF IMPURE STATES}

In classical statistical physics a
mixture  state is interpreted as a statistical average over an ensemble in
which any individual realizations is in a pure state.
In  quantum physics
a mixture can be considered  a state of a
quantum system, which  can not be completely
described in terms of its own Hilbert space, because it is only a part
of a more complex quantum system. Due to the lack of information about
other parts of this complex system, the description of subsystem is possible
only in terms of mixtures.

Let assume quantum system $P$ being entangled with some other quantum system
$R$ (reservoir). Let the composed system $S$ (composed of $P$ and $R$) itself
is in a pure state $\vert \Psi\rangle$. The density operator $\hat{\rho}_P$ of
the system $P$ is then obtained via tracing over the reservoir degrees of
freedom:
\begin{eqnarray}
\hat{\rho}_P={\rm Tr}_R\,\left[\hat{\rho}_S\right];\qquad
\hat{\rho}_S =|\Psi\rangle\langle \Psi| .
\label{10.1}
\end{eqnarray}
Once the system $S$ is in a pure state, then we can safely apply
the Bayesian reconstruction scheme as described in Section 2.
The reconstruction itself is based only on data associated with
measurements performed on the system $P$. When the density operator
$\hat{\rho}_S$ is {\em a posteriori} estimated, then by tracing
over the reservoir degrees of freedom, we obtain the {\em a posterior}
density operator $\hat{\rho}_P$ for the system $P$ (with no {\em a priori}
constraint on the purity of the state of the system $P$). These arguments
are equivalent to the ``purification'' Ansatz as proposed by Uhlmann
\cite{Uhlmann}.

To make our reconstruction scheme for impure state consistent
we have to chose the reservoir $R$ uniquely. This can be
done with the help of the Schmidt theorem (see Ref.\cite{Ekert95})
from which it follows that if the composite system $S$ is in a pure
state $|\Psi\rangle$ then its state vector can be written in the form:
\begin{eqnarray}
|\Psi\rangle=\sum_{i=1}^M c_i \vert\alpha_i\rangle_{_P}\otimes\vert
\beta_i\rangle_{_R},
\label{10.2}
\end{eqnarray}
where $\vert\alpha_i\rangle_{_P}$ and $\vert\beta_i\rangle_{_R}$ are
elements from two specific
orthonormalized bases associated with the subsystems $P$ and $R$,
respectively,
and $c_i$ are appropriate complex  numbers satisfying the
normalization condition  $\sum \vert c_i \vert^2=1$.
The maximal index of summation ($M$) in Eq.(\ref{10.2}) is given by the
dimensionality of the Hilbert space of the system $P$.
In other words, when we apply the Bayesian method  to the case of impure
states
of $M$-level system, it is sufficient to ``couple'' this system  to an
$M$-dimesional ``reservoir''.   Due to the fact that
we measure only  observables of the first subsystem $P$ particular form of
states	$\vert\beta_i\rangle_{_R}$ of
the second subsystem $R$  does not affect  results of the reconstruction.

\section{SPIN-1/2 RECONSTRUCTION}

We assume an ensemble of
spins-$1/2$ in an unknown state described by the density operator
\begin{eqnarray}
\hat\rho(\theta,\phi)= {1\over 2}\left( \hat{1} + \vec{r}\,
\hat{\vec{\sigma}}
\right)
\label{5.1}
\end{eqnarray}
where
$\vec{r}=(\sin \theta \cos \phi,
\sin \theta \sin \phi, \cos \theta)$;
 $\phi \in (0,2 \pi),\ \theta \in (0,\pi)$,
and $\hat{1}$ is the unity operator.
The Pauli spin operators $\vec{\hat{\sigma}}$
 in the matrix representation in the basis
$|0\rangle$, $|1\rangle$ of the eigenvectors of the operator $\sigma_z$
do read
\begin{eqnarray}
\hat{\sigma}_x=\left(\begin{array}{cc} 0 & 1 \\ 1 & 0 \end{array}\right),
\quad
\hat{\sigma}_y=\left(\begin{array}{cc} 0 & -i \\ i & 0 \end{array}\right),
\quad
\hat{\sigma}_z=\left(\begin{array}{cc} 1 & 0 \\ 0 & -1 \end{array}\right)
.\label{5.2}
\end{eqnarray}

To determine completely the unknown state, one has to measure
three linearly independent (e.g., orthogonal) projections of the spin-1/2.
One possible choice of	the complete set of
observables (i.e., the quorum \cite{Band79})
associated with the spin-1/2 are  spin projections for three
orthogonal directions represented by the Hermitian operators:
\begin{eqnarray}
\hat{s}_i\equiv {{\hat{\sigma}_i}\over2},\qquad i=x,y,z
\label{5.3}
\end{eqnarray}

In what follows we will consider
 three observation levels defined as
${\cal O}_A=\{\hat s_z \}$,
${\cal O}_B=\{ \hat s_z,\hat s_x \}$ and
${\cal O}_C=\{\hat s_z,\hat s_x,\hat s_y \}
\equiv{\cal O}_{comp}$.

In order to apply the Bayesian inference scheme for impure states, we have
to ``purify'' \cite{Uhlmann} the quantum-mechanical system as discussed
in Section IV. In the particular case of the spin-1/2 it means that
we have to consider a system of two spins-1/2 on of which does play
the role of a reservoir.  In this case the corresponding two-spins-1/2
density
operator can be parameterized as
\begin{eqnarray}
\hat{\rho}(\alpha,\psi,\phi_1,\theta_1,\phi_2,\theta_2)={\hat 1\otimes\hat 1
\over4}
+{{\vec r^{_{(1)}}\hat{\vec{\sigma}}\otimes\vec r^{_{(2)}}
\hat{\vec{\sigma}}} \over4}
+\cos\alpha\Big\lbrack {{\vec r^{_{(1)}}\hat{\vec{\sigma}}
\otimes\hat 1}\over4}+
{{\hat1\otimes\vec r^{_{(2)}}\hat{\vec{\sigma}}}\over4}\Big\rbrack
\label{8.5}
\end{eqnarray}
\begin{eqnarray}\nonumber
+\sin\alpha\cos\psi\Big\lbrack {{{\vec k^{_{(1)}}\hat{\vec{\sigma}}}\otimes
\vec k^{_{(2)}}\hat{\vec{\sigma}} }\over4}
-{{\vec l^{_{(1)}}\hat{\vec{\sigma}}\otimes\vec l^{_{(2)}}\hat{\vec{\sigma}}
 }\over4}\Big\rbrack
-\sin\alpha\sin\psi\Big\lbrack {{\vec k^{_{(1)}}\hat {\vec{\sigma}}\otimes
\vec l^{_{(2)}}\hat{\vec{\sigma}} }\over4}
+{{\vec l^{_{(1)}}\hat{\vec{\sigma}}\otimes\vec k^{_{(2)}}\hat{\vec{\sigma}}
 }\over4}\Big\rbrack,
\end{eqnarray}
where $\psi,\phi_1,\phi_2 \in (0,2\pi)$;
$\alpha, \theta _1,\theta _2 \in (0,\pi)$ and
\begin{eqnarray}
\vec k^{_{(j)}}&=&(\sin \phi_j,-\cos \phi_j,0);\nonumber \\
\vec l^{_{(j)}}&=&(\cos\theta_j\cos\phi _j,\cos\theta_j\sin\phi _j,
-\sin\theta_j);
\label{8.6}\\
\vec r^{_{(j)}}&=&(\sin\theta_j\cos\phi_j,\sin\theta_j\sin\phi_j,
\cos\theta_j).\nonumber
\end{eqnarray}
Once we have parameterized the state space $\Omega$ we have to find the
invariant integration measure $d_{\Omega}$. We have derived this measure
earlier \cite{Derka96}
and it reads:
\begin{eqnarray}
d_{\Omega}=\cos^2\alpha \sin\alpha \sin\theta_1 \sin\theta_2 d\alpha
d\psi d\phi_1
d\theta_1 d\phi_2 d\theta_2.
\label{8.15}
\end{eqnarray}

A set of projectors associated with
the observables $\hat{\vec{\sigma}}^{_{(1)}}$ and
$\hat{\vec{\sigma}}^{_{(2)}}$
[in what follows we use the notation such that the position of the operator
to the left (right) of the symbol $\otimes$ is associated with the
first (second) spin-1/2]:
\begin{eqnarray}
\hat P_{s,\hat{s}_i^{_{(1)}}}={({\hat 1+s\hat{\sigma}_i)}\over 2}
\otimes \hat 1 ;\qquad
\hat P_{s,\hat{s}_i^{_{(2)}}}=\hat 1 \otimes {({\hat 1+s\hat{\sigma}_i)}
\over 2}; \qquad
\hat P_{s,\hat{s}_i^{_{(1)}}\hat{s}_j^{_{(2)}}}={\hat 1 \otimes \hat 1
\over 2}+
s{\hat{\sigma}_i\otimes \hat {\sigma}_j\over 2}.
\label{9.1}
\end{eqnarray}
The corresponding conditional probabilities do read
\begin{eqnarray}
p(s,\hat{s}_i^{_{(1)}} \vert \hat{\rho} (\alpha \dots ))=
{1 \over 2}+s{{\cos(\alpha)}\over 2}r_{i}^{_{(1)}};\qquad
p(s,\hat{s}_i^{_{(2)}} \vert \hat{\rho} (\alpha \dots ))=
{1 \over 2}+s{{\cos(\alpha)}\over 2}r_{i}^{_{(2)}};
\label{9.2}
\end{eqnarray}
\begin{eqnarray}\nonumber
p(s,\hat{s}_i^{_{(1)}}\hat{s}_j^{_{(2)}} \vert \hat{\rho} (\alpha \dots ))=
{1 \over 2}+s{{r_i^{_{(1)}}r_j^{(2}}\over2}+
s\frac{\sin(\alpha)}{2}
\left[(k_i^{_{(1)}}k_j^{_{(2)}}-l_i^{_{(1)}}l_j^{_{(2)}})\cos\psi
-(k_i^{_{(1)}}l_j^{_{(2)}}+l_i^{_{(1)}}k_j^{_{(2)}})\sin\psi  \right].
\end{eqnarray}

We remind ourselves that we do consider only measurements performed on
the first
spin described by the observables $\hat{\vec{\sigma}}^{_{(1)}}$. After the
Bayesian reconstruction of the composed system is performed then the
``reservoir'' degrees of freedom are traced out. The resulting density
operator describes an {\em a posteriori} estimation of the density
operator of a two-level system with no reference on the {\em a priori}
assumption about the purity of the spin.

Instead of analysing estimated density operators after a finite number
of measurements performed over the spin-1/2 (i.e. measurements performed
over a finite number of elements of the ensemble) we focus our attention
on results of the reconstruction in the limit of infinite number
of measurements.

\subsection{Observation level ${\cal O}_A=\{\hat s_z \}$}
On the observation level ${\cal O}_A$
only the spin  component $\hat s_z $ is measured.
This kind of the measurement can be performed with the
help of one Stern-Gerlach apparatus.
With the help of the data obtained in
a large (infinite) number of measurements we can  express
the density operator of the spin-1/2 under consideration as
(the trace over the reservoir degrees of freedom has already been performed)
\begin{eqnarray}
\hat\rho=\frac{1}{\cal N}
\int_{-1}^{1}y^2dy\int_0^\pi\sin\theta_1 d\theta_1
\ \ \delta(\langle \hat \sigma_z^{_{(1)}} \rangle -y\cos\theta_1)
(\hat 1+y\cos\theta_1\hat\sigma_z),
\label{11.1}
\end{eqnarray}
where the variable $\alpha$ is substituted by $y=\cos\alpha$.
When we perform  integration  over the variable $y$ we obtain the expression
\begin{eqnarray}
\hat\rho=\frac{1}{\cal N} \int_{{\cal L}'}
  d\theta_1
{\sin\theta_1\over \cos^2\theta_1\vert \cos\theta_1\vert}
(\hat 1 +\langle \hat \sigma_z^{_{(1)}} \rangle \hat\sigma_z),
\label{11.2}
\end{eqnarray}
where the integration is performed over the region
\begin{eqnarray}
{\cal L}' := \{0,\pi\}~~~\mbox{ such that }~~
\vert \cos\theta_1 \vert \geq \vert \langle\hat \sigma_z^{_{(1)}}\rangle\vert.
\label{11.3}
\end{eqnarray}
After we perform the integration over $\theta_1$ we obtain
for the {\em a posteriori} estimation of the density operator the
expression
\begin{eqnarray}
\hat \rho =
{1 \over 2}(\hat 1 +\langle \hat \sigma_z \rangle \hat \sigma_z).
\label{5.7}
\end{eqnarray}
which is identical to the one obtained via the Jaynes principle of
maximum entropy \cite{Drobny96}.

\subsection{Observation level ${\cal O}_B=\{ \hat s_z,\hat s_x \}$}

Let us extend the observation level ${\cal O}_A$ and let us assume
the measurement of two
spin projections $\hat s_z$ and $\hat s_x$ (i.e. two Stern-Gerlach
apparatuses with fixed orientations are employed).
In the limit of infinite number of measurements one can express
the Bayesian estimation  of the
density operator of the spin-1/2
on the given observation level as (here the trace over the ``reservoir''
spin has already been performed):
\begin{eqnarray}\nonumber
\hat\rho =\frac{1}{\cal N}
\int_{-1}^{1}\!\!y^2dy\!\!\int_0^\pi\!\!\sin\theta_1 d\theta_1
\!\!\int_0^{2\pi}\!\!d\phi_1
\ \ \delta(\langle \hat \sigma_z^{_{(1)}} \rangle -y\cos\theta_1)
\delta(\langle \hat \sigma_x^{_{(1)}} \rangle -y\sin\theta_1\cos \phi_1)
\end{eqnarray}
\begin{eqnarray}
\times(\hat 1+y\sin\theta_1\cos \phi_1\hat\sigma_x+y\sin\theta_1\sin
\phi_1\hat\sigma_y+
y\cos\theta_1\hat\sigma_z).
\label{11.4}
\end{eqnarray}
When we perform integration over the variable $y$ we find
\begin{eqnarray}\nonumber
\hat\rho=\frac{1}{\cal N}\int_0^{2\pi}d\phi_1
\int_ {{\cal L}'}
d\theta_1
{\sin\theta_1\over \cos^2\theta_1\vert \cos\theta_1\vert}
 \delta(\langle \hat \sigma_x^{_{(1)}} \rangle -\tan\theta_1\cos \phi_1
\langle \hat \sigma_z^{_{(1)}}\rangle)
\end{eqnarray}
\begin{eqnarray}
\times(\hat 1 +\langle \hat \sigma_z^{_{(1)}}\rangle \tan\theta_1
\cos\phi_1\hat\sigma_x+
\langle \hat \sigma_z^{_{(1)}} \rangle \tan\theta_1\sin \phi_1\hat\sigma_y+
\langle \hat \sigma_z^{_{(1)}} \rangle \hat\sigma_z).
\label{11.5}
\end{eqnarray}
The integration over the variable $\phi_1$ in the right-hand side of
Eq.(\ref{11.5}) results into the following expression
\begin{eqnarray}
\hat\rho=\frac{1}{\cal N} \int_{{\cal L}''}
\, d\theta_1
\sum_{j=1}^2
{1 \over
\cos^2\theta_1\vert\sin\phi_{1}^{(j)}\vert}
(\hat 1 +\langle \hat \sigma_x^{_{(1)}}\rangle \hat\sigma_x+
\langle \hat \sigma_z^{_{(1)}} \rangle \tan\theta_1\sin \phi_{1}^{(j)}
\hat\sigma_y+
\langle \hat \sigma_z^{_{(1)}} \rangle \hat\sigma_z),
\label{11.6}
\end{eqnarray}
where the integration is performed over the region
\begin{eqnarray}
{\cal L}'' := \{0,\pi\}~~~\mbox{ such that }~~
\vert \cos\theta_1 \vert \geq \vert \langle\hat \sigma_z^{_{(1)}}\rangle\vert,
~~~\mbox{and}~~~
\vert \tan \theta_1 \vert \geq \left\vert \frac{\langle \hat \sigma_x^{_{(1)}}
\rangle}{
\langle \hat \sigma_z^{_{(1)}} \rangle }\right\vert.
\label{11.7}
\end{eqnarray}
The sum in Eq.(\ref{11.6}) is performed over two values $\phi_1^{(j)}$
of the variable $\phi_1$ which are equal to  two solutions
of the equation
\begin{eqnarray}
\cos \phi_{1}=\frac{\langle \hat \sigma_x^{_{(1)}} \rangle}{
\langle \hat \sigma_z^{_{(1)}} \rangle \tan \theta_1}.
\label{11.8}
\end{eqnarray}
Due to the fact that the term in front of the operator
$\hat{\sigma}_y^{_{(1)}}$
is an odd function of $\phi_1^{(j)}$, we can straightforwardly perform
integration over $\theta_1$ and we find the expression for the reconstructed
density operator
\begin{eqnarray}
\hat \rho =
{1\over 2}(\hat 1 + \langle \hat \sigma_z \rangle \hat \sigma_z
 + \langle \hat \sigma_x \rangle \hat \sigma_x),
\label{5.10}
\end{eqnarray}
which again is exactly the same as if we perform
the reconstruction with the help of the Jaynes principle
\cite{Drobny96}.

\subsection{Complete observation level	${\cal O}_C=\{\hat s_z^{_{(1)}},
\hat s_x^{_{(1)}} ,
\hat s_y^{_{(1)}} \}$}
Let us assume now that the measurement is performed with the help
of three Stern--Gerlach apparatuses, each of which are measuring
three spin component $s_i^{(1)}$ ($i=x,y,z$).
On this complete observation level the expression for the Bayesian
estimation of the density operator of the spin-1/2 in the
limit of infinite number of measurements can be expressed as
(here again we have already traced over the ``reservoir'' degrees
of freedoms):
\begin{eqnarray}\nonumber
\hat\rho=\frac{1}{\cal N}
\int_{-1}^{1}\!\!y^2dy\!\!\int_0^\pi\!\!\sin\theta_1 d\theta_1
\!\!\int_0^{2\pi}\!\!d\phi_1
 \delta(\langle \hat \sigma_z^{_{(1)}} \rangle -y\cos\theta_1)
\delta(\langle \hat \sigma_x^{_{(1)}} \rangle -y\sin\theta_1\cos \phi_1)
\end{eqnarray}
\begin{eqnarray}
\times
\delta(\langle \hat \sigma_y^{_{(1)}} \rangle -y\sin\theta_1\sin \phi_1)
(\hat 1+y\sin\theta_1\cos \phi_1\hat\sigma_x+y\sin\theta_1\sin \phi_1
\hat\sigma_y+
y\cos\theta_1\hat\sigma_z).
\label{11.9}
\end{eqnarray}
We  can rewrite Eq.(\ref{11.9}) as
\begin{eqnarray}\nonumber
\hat\rho=\frac{1}{\cal N}
\int_{{\cal L}''}
\, d\theta_1
\sum_{j=1}^2
{1 \over
\cos^2\theta_1\vert\sin\phi_{1}^{(j)}\vert}
 \delta\left(\langle \hat \sigma_y^{_{(1)}}\rangle -\tan\theta_1
\sin \phi_{1}^{(j)}
\langle \hat \sigma_z^{_{(1)}}\rangle \right)
\end{eqnarray}
\begin{eqnarray}
\times
\left[\hat 1 +\langle \hat \sigma_x^{_{(1)}}\rangle \hat\sigma_x+
\langle \hat \sigma_y^{_{(1)}}\rangle \tan\theta_1\sin \phi_{1}^{(j)}
\hat\sigma_y+
\langle \hat \sigma_z^{_{(1)}}\rangle \hat\sigma_z \right],
\label{11.10}
\end{eqnarray}
where ${\cal L}''$ and $\phi_1^{(j)}$ are defined by Eqs.(\ref{11.7})
and (\ref{11.8}), respectively. Now the
integration over parameter $\theta_1$ can be easily performed and for the
density operator of the given spin-1/2 system we find
\begin{eqnarray}
\hat\rho=\frac{1}{2}
(\hat 1 + \langle \hat \sigma_x^{_{(1)}}\rangle \hat\sigma_x+
\langle \hat \sigma_y^{_{(1)}}\rangle \hat\sigma_y+
\langle \hat \sigma_z^{_{(1)}}\rangle \hat\sigma_z).
\label{11.11}
\end{eqnarray}
The density operator  (\ref{11.11}) obtained with the help  of Bayesian
inference scheme is equal to that one  which follows from the Jaynes
principle of the maximum entropy. Moreover this same result can be obtained
from other reconstruction schemes, such as the discrete quantum tomography
\cite{Leon95b} (see also \cite{Band79} and \cite{Newton68}).

The simple example of the spin-1/2 reconstruction
reveals deep conceptual relationship between the quantum Bayesian inference and
the Jaynes principle of the maximum entropy. To understand this relationship
more clearly we turn our attention once again to the {\em a priori} assumption
under which we have performed the reconstruction. We have assumed that the
spin-1/2 can be in an impure state (therefore we have applied
the ``purification'' procedure).
As a consequence of this assumption the von\,Neumann entropy of
the reconstructed density operator may be larger than zero, which
is equivalent to the fact that the mean values of the observables
$\hat{\sigma}_i$ ($i=x,y,z$) do fulfill the condition
\begin{eqnarray}
\langle\hat{\sigma}_x\rangle^2+
\langle\hat{\sigma}_y\rangle^2+
\langle\hat{\sigma}_z\rangle^2 \leq 1,
\label{5.15}
\end{eqnarray}
i.e. the reconstructed state can be expressed as a point either on or
inside the Poincare sphere.

On the contrary, if it is {\em a priori} assumed that the reconstructed
state is a pure one (see, for instance, works by Jones \cite{Jones91}),
then the Bayesian reconstruction in the limit of infinite number of
measurements
on the complete observation level
results in the
reconstructed density operator which can be expressed as
\begin{eqnarray}\nonumber
\hat\rho =\frac{1}{\cal N}
\int_0^{2\pi}\!\!d\phi\!\!\int_0^\pi\!\!\sin\theta d\theta
\ \ \delta(\langle \hat \sigma_z \rangle -\cos\theta)
\delta(\langle \hat \sigma_x \rangle -\sin\theta\cos\phi)
\delta(\langle \hat \sigma_y \rangle -\sin\theta\sin\phi)
\end{eqnarray}
\begin{eqnarray}
\times(\hat 1 + \sin\theta\cos\phi
\hat\sigma_x + \sin\theta\sin\phi\hat\sigma_y +
\cos\theta\hat\sigma_z).
\label{6.11}
\end{eqnarray}
The integral in the right-hand side of the equation can only be performed
if the mean values of the observables under consideration do fulfill
the purity condition
\begin{eqnarray}
\langle\hat{\sigma}_x\rangle^2+
\langle\hat{\sigma}_y\rangle^2+
\langle\hat{\sigma}_z\rangle^2 =1.
\label{5.16}
\end{eqnarray}
Providing the purity condition holds then from Eq.(\ref{6.11}) we obtain
for the density operator the expression (\ref{11.11}), otherwise the
reconstruction scheme fails.

The limit formulae  for the Bayes inference
 have an appealing geometrical interpretation. For example the three
$\delta$-functions in Eq.(\ref{6.11}) correspond to three specific
orbits on the
Poincare sphere, each of which is associated with a set of pure states
which posses the measured value of a given observable $\hat{s}_i$.
This corresponds to an averaging over the {\em generalized microcanonical}
ensemble of pure states having the measured mean values of the three
observables.
The reconstructed density operator then describes a point on the Poincare
sphere which coincides with an intersection of these three ``orbits''.
Consequently, if the three orbits have no intersection, the reconstruction
scheme fails, because
there does not exist a {\em pure} state with the given mean values
of the measured observables.

On the other hand, the form of the {\em a posteriori} expression for
the reconstructed density operator (\ref{11.9})  obtained with the
help of the Bayesian inference with no {\em a priori} restriction
on the purity of the reconstructed state reveals that the estimated
density operator can be obtained as a result of averaging over all
points on and inside the Poincare sphere. This
averaging over the {\em generalized canonical ensemble} is
equal to the maximization  of the von\,Neumann
entropy  as assumed by Jaynes.

\section{CONCLUSIONS}

In the paper
we have analyzed in detail the logical connection between three different
reconstruction schemes: {\bf (1)} If  measurements over a finite
number of elements of the ensemble are performed then one
can obtain the {\em a posteriori} estimation of the density operator
with the help of the
Bayesian inference. If nothing is know about the reconstructed state
one has to
assume a constant prior probability distribution on the parametric state space
 under the assumption that the system is in a
statistical mixture.  {\bf (2)} As soon as number of measurements becomes
large then
the Bayesian inference scheme becomes equal to the reconstruction scheme
based on the Jaynes principle of the maximum entropy, i.e.,
 in the limit of infinite number
of measurements {\em a posteriori} estimated density operator fulfills
the condition of the maximum entropy. Consequently, it is equal to the
generalized canonical density operator. {\bf (3)} If the quorum of observables
is measured, then the generalized canonical operator is equal to the
``true'' density operator of the system itself, i.e. the complete
reconstruction via the MaxEnt principle is performed. It is the question
of technical convenience which reconstruction scheme on the complete
observation level is utilized (for instance,  quantum tomography
can be used),
 but the fact is that all of them
can be formulated as a maximization of the entropy under given constraints.

\vspace{0.5truecm}

{\bf Acknowledgements}\newline
This work was in part supported by the United Kingdom Engineering
and Physical Sciences Research Council, the European Community,
and the Grant Agency
VEGA of the Slovak Academy of Sciences (grant. n. 2/1154/96).
We acknowledge the support by the  East-West Program of the Austrian
Academy of Sciences under the contract No. 45.367/6-IV/3a/95 of the
\"{O}sterreichisches Bundesministerium f\"{u}r Wissenschaft und Forschung.


\begin{references}

\bibitem{Ballentine90}
{ L.E. Ballentine}: {\em Quantum Mechanics}
(Prentice Hall, Englewood Cliffs, New Jersey, 1990).

\bibitem{Band79}
 W. Band and J.L. Park,
  {\em Am. J. Phys.} {\bf 47}, 188 (1979);
  {\em Found. Phys.} {\bf 1}, 133 (1970);
  {\em Found. Phys.} {\bf 1}, 339 (1971);
  J.L. Park and W. Band,
  {\em Found. Phys.} {\bf 1}, 211 (1971).

\bibitem{Neumann55}
{  J. von\,Neumann}: {\em
  Mathematical Foundations of Quantum Mechanics}
  (Princeton University Press, Princeton, 1955) see also
{ J.A. Wheeler} and { W.H. Zurek}:
  {\em Quantum Theory and Measurement}
  (Princeton University Press, Princeton, 1983).


\bibitem{Pauli80}
{ W. Pauli}:
  {\em General Principles of Quantum Mechanics}
  (Springer Verlag, Berlin, 1980); see also
{B. d'\,Espagnat}:
  {\em Conceptual Foundations of Quantum Mechanics}, 2$^{\rm nd}$ ed.
  (W.A.Benjamin, Reading, 1976).

\bibitem{Gale68}
{ W. Gale, E. Guth}, and {G.T. Trammel},
{\em Phys. Rev.} {\bf 165}, 1434 (1968); see also
{ A. Orlowski} and { H. Paul},
{\em Phys. Rev.} {\bf 50}, R921 (1994).


\bibitem{Vogel89}
{  K.Vogel} and {  H.Risken},
  {\em Phys. Rev. A} {\bf 40}, 2847 (1989); see also
{  J.Bertrand} and {  P.Bertrand},
  {\em Found. Phys.}  {\bf 17}, 397 (1987);
{  M.Freyberger, K.Vogel,} and {  W.P.Schleich},
  {\em Phys. Lett. A} {\bf 176}, 41 (1993);
{  H.K\"{u}hn, D.-G.Welsch}, and {  W.Vogel},
  {\em J. Mod. Opt.} {\bf 41}, 1607 (1994);
{  G.S.Agarwal} and {  S.Chaturvedi},
  {\em Phys. Rev. A} {\bf 49}, R665 (1994);
{  W.Vogel} and {  D.-G.Welsch},
  {\em Acta Phys. Slov.} {\bf 45}, 313 (1995);
 U. Leonhardt, M. Munroe, T. Kiss, T. Richter, and M.G. Raymer,
  {\em Opt. Commun.} {\bf 127}, 144 (1996);
 U. Leonhardt, H. Paul, and G.M. D'Ariano,
 {\em Phys.  Rev. A} {\bf 52}, 4899 (1995);
 G.M. D'Ariano, U. Leonhardt, and H. Paul,
 {\em Phys. Rev. A} {\bf 52}, R1801 (1995);
 U. Leonhardt and H. Paul,
 {\em Phys. Lett. A} {\bf 193}, 117 (1994);
 {\em J. Mod. Opt.} {\bf 41}, 1427 (1994).

\bibitem{Smithey93}
{  D.T. Smithey, M. Beck, M.G. Raymer} and {  A. Faridani},
  {\em Phys. Rev. Lett.}{\bf 70}, 1244 (1993);
{  D.T. Smithey, M. Beck, J. Cooper} and {  M.G. Raymer},
  {\em Phys. Scr. T} {\bf 48}, 35 (1993);
  M. Beck, M.G. Raymer, I.A. Wamsley, and V. Wong,
 {\em Opt. Lett.} {\bf 18}, 2041 (1993).
  M. Beck, D.T. Smithey, and M.G. Raymer,
  {\em Phys. Rev. A} {\bf 48}, R890 (1993);
see also
{  M.G. Raymer, M. Beck}, and {  D.F. Mc\,Alister},
  {\em Phys. Rev. Lett.} {\bf 72}, 1137 (1994);
{  M.G. Raymer, D.T. Smithey, M. Beck, and J. Cooper},
 {\em Acta Phys. Pol.} {\bf 86}, 71 (1994).


\bibitem{Wilkens}
{ U. Janicke and M. Wilkens}, {\em J. Mod. Opt.} {\bf 42}, 2183 (1995).

\bibitem{Leon95b}
{  U.Leonhardt}, {\em Phys.Rev.Lett.} {\bf 74}, 4101 (1995);
{\em Phys. Rev. A} {\bf 53}, 2998 (1996); see also
{  W.K.Wootters}, {\em Found. Phys.} {\bf 16}, 391 (1986).
{ W.K.Wootters}, {\em Ann. Phys. (N.Y.)} {\bf 175}, 1 (1987);
{D.Galetti} and { A.F.R. De\,Toledo\,Piza},
  {\em Physica A} {\bf 149}, 267 (1988).

\bibitem{Husimi}
{ K. Husimi},
  {\em Proc. Phys. Math. Soc. Jpn.} {\bf 22},  264 (1940);
{ Y. Kano},
  {\em J. Math. Phys.} {\bf 6}, 1913 (1965).

\bibitem{Arthurs65}
{  E. Arthurs} and {  J.L. Kelly, Jr.},
  {\em Bell. Syst. Tech. J.} {\bf 44}, 725 (1965);
{ W.K. Wootters} and { W.H. Zurek},
  {\em Phys. Rev. D} {\bf 19}, 473 (1979);
{  Y.Lai} and {  H.A.Haus},
  {\em Quantum Opt.} {\bf 1}, 99 (1989);
{  D.Lalovi\'c, D.M.Davidovi\'c}, and {  N.Bijedi\'c},
  {\em Phys. Rev. A} {\bf 46}, 1206 (1992);
{  D.M.Davidovi\'c} and {  D.Lalovi\'c},
  {\em J. Phys. A} {\bf 26}, 5099 (1993);
{  S.Chaturvedi, G.S.Agarwal}, and {  V.Srinivasan},
  {\em J. Phys.A} {\bf 27},  L39 (1994);
{  M.G.Raymer},
{\em Am. J. Phys.} {\bf 62}, 986 (1994);
 U. Leonhardt and H. Paul,
{\em Phys. Rev. A} {\bf 48}, 4598 (1993);
V. Bu\v{z}ek, C.H. Keitel, and P.L. Knight, {\em Phys. Rev. A} {\bf 51},
2575 (1995);
see also the review articles
{  S.Stenholm},
  {\em Ann. Phys. (N.Y.)} {\bf 218}, 233 (1992), and
U.  Leonhardt and H. Paul, {\em Quant. Electron.}  {\bf 19}, 89 (1995).

\bibitem{Wod84}
{K. W\'odkiewicz}
  {\em Phys. Rev. Lett.} {\bf 52}, 1064 (1984);
  {\em Phys. Lett. A} {\bf 115}, 304 (1986);
  {\em Phys. Lett. A} {\bf 129}, 1 (1988).

\bibitem{Walker86}
N.G. Walker and J.E. Carroll, {\em Opt. Quant. Electron.} {\bf 18}, 335 (1986);
{ N.G. Walker},
  {\em J. Mod. Opt.} {\bf 34}, 15 (1987).

\bibitem{Noh91}
{ J.W. Noh, A. Foug\'eres}, and { L. Mandel},
  {\em Phys. Rev. Lett.} {\bf 67}, 1426 (1991);
  {\em Phys. Rev. A} {\bf 45}, 424 (1992).

\bibitem{Perina}
{ J. Pe\v{r}ina, Z. Hradil}, and { B. Jur\v{c}o}:
  {\em Quantum Optics and Fundamentals of Physics} (Kluwer
  Academic Publishers, Dordrecht, 1994).

\bibitem{Opat95}
{T. Opatrn\'y, V. Bu\v{z}ek, J. Bajer}, and {G. Drobn\'y},
   {\em Phys. Rev. A} {\bf 52}, 2419 (1995);
{T. Opatrn\'y, D.-G. Welsch}, and {V. Bu\v{z}ek},
 {\em Phys. Rev. A} {\bf 53}, 3822 (1996).


\bibitem{Jaynes57}
E.T. Jaynes, {\em Phys. Rev.} {\bf 108}, 171 (1957);
{\em ibid.~} {\bf 108}, 620 (1957) 620;
{\em Am. J. Phys.} {\bf 31}, 66 (1963).
 E.T. Jaynes,
  ``Information theory and  statistical mechanics'',
  in {\em 1962 Brandeis Lectures, Vol 3},
  ed. K.W. Ford (Benjamin, Inc. New York, 1963), p.181.
See also
 E. Fick and G. Sauermann:
{\em  The Quantum Statistics of Dynamic Processes}
(Springer Verlag, Berlin, 1990);
J.N.Kapur and H.K.Kesavan:
{\em Entropic Optimization Principles with Applications}
(Academic Press, New York, 1992).


\bibitem{comment}
The {\em MaxEnt} principle has a very close
relation to the Laplace principle of indifference [see, for
instance, H. Jeffreys: {\em Theory of Probability}
(Oxford Univ. Press., Oxford, 1960)] which states that where nothing
is known one should choose a constant-valued function to reflect this
ignorance. This obviously maximizes any uncertainty measure.



\bibitem{Buzek96}
V. Bu\v{z}ek, G. Adam, and G. Drobn\'y, {\em Ann. Phys. (N.Y.)} {\bf 245},
37 (1996).

\bibitem{Wiedemann96}
H. Wiedemann: {\em Quantum tomography with the maximum entropy principle}
(unpublished).


\bibitem{Helstrom76}
C.W. Helstrom: {\em Quantum Detection and Estimation Theory}
(Academic Press, New York, 1976).

\bibitem{Holevo82}
A.S. Holevo: {\em Probabilistic and Statistical Aspects of Quantum
Theory} (North-Holland, Amsterdam, 1982).

\bibitem{Jones91}
K.R.W. Jones, {\em Ann. Phys. (N.Y.)} {\bf 207}, 140 (1991);
{\em Phys. Rev. A} {\bf 50}, 3682 (1994).  For more discussion see
Z. Hradil: {\em Quantum state estimation}, Los Alamos e-print
archive, quant-ph/9609012.


\bibitem{Derka96}
R. Derka, V. Bu\v{z}ek, and G. Adam, {\em Acta Phys. Slov.}
{\bf 46}, 355 (1996); see also
R. Derka, V. Bu\v{z}ek, G. Adam,  and P.L. Knight
(unpublished).

\bibitem{Gradstein}
I.S. Gradstein and I.M. Ryzhik: {\em Table of Integrals, Series, and
Products} (Academic Press, New York, 1980).


\bibitem{Uhlmann}
A. Uhlmann,
   {\em Rep. Math. Phys.} {\bf 9}, 273 (1976);
   {\em ibid} {\bf 24}, 229 (1986).

\bibitem{Ekert95}
A.K. Ekert and P.L. Knight, {\em Am. J. Phys.} {\bf 63}, 415 (1995).


\bibitem{Drobny96}
{ G. Drobn\'{y}, R. Derka, G. Adam, and V. Bu\v{z}ek}:
``{\em Reconstruction of quantum states of spin systems via the
Jaynes principle of maximum entropy}'' to appear in the
special issue of {\em J. Mod. Opt.} (1997).

\bibitem{Newton68}
{  R.G.Newton} and {  Bing-Lin Young},
  {\em Ann. Phys.} (N.Y.) {\bf 49}, 393 (1968).


\end{references}
\end{document}